\documentclass[ reprint,
superscriptaddress,
bibnotes,
 amsmath,amssymb,
 aps,
pra,
]{revtex4-2}
\usepackage{lipsum}
\usepackage{graphicx}
\usepackage{soul} 
\usepackage{dcolumn}
\usepackage{bm}
\usepackage[caption=false]{subfig} 
\usepackage{cancel}
\usepackage{comment}
\usepackage{hyperref}
\usepackage[usenames, dvipsnames]{color}
\usepackage{hyperref}
\hypersetup{
    colorlinks=true,
    linkcolor=Blue,
    citecolor=Blue,
    filecolor=Blue,      
    urlcolor=Blue,
    }

\newcommand{\sech}{\textrm{sech}}

\newcommand{\stfirst}[1]{}%
\newcommand{\stsecond}[1]{}%
\newcommand{\red}[1]{{\color{red}#1}}%
\newcommand{\redsecond}[1]{{\color{red}#1}}%
\renewcommand{\red}[1]{#1}%
\renewcommand{\redsecond}[1]{#1}%

\begin{document}

\preprint{APS/123-QED}

\title{Variational Approach to the Snake Instability of a Bose-Einstein Condensate Soliton}
\author{U. Tanyeri}
 \email{tanyeri@metu.edu.tr}
 \affiliation{%
Department of Physics, Middle East Technical University, Ankara, 06800, Turkey
}%
\author{M.A. G\"urkan}
 \affiliation{%
Department of Physics, Middle East Technical University, Ankara, 06800, Turkey
}%
\author{A. Kele\c{s}}
 \affiliation{%
Department of Physics, Middle East Technical University, Ankara, 06800, Turkey
}%
\author{M.\"O. Oktel}
\affiliation{Department of Physics, Bilkent University, Ankara, 06800, Turkey}

\date{\today}

\begin{abstract}

Solitons are striking manifestations of nonlinearity, encountered in diverse physical systems such as water waves, nonlinear optics, and Bose-Einstein condensates (BECs). In BECs, dark solitons emerge as exact stationary solutions of the one-dimensional Gross-Pitaevskii equation. While they can be long-lived in elongated traps, their stability is compromised in higher dimensions due to the snake instability, which leads to the decay of the soliton into vortex structures among other excitations. We investigate the dynamics of a dark soliton in a Bose-Einstein condensate confined in an anisotropic harmonic trap. Using a variational ansatz that incorporates both the transverse bending of the soliton plane and the emergence of vortices along the nodal line, we derive equations of motion governing the soliton's evolution. This approach allows us to identify stable \red{soliton-vortex} oscillation modes as well as the growth rates of the unstable perturbations. In particular, we determine the critical trap anisotropy required to suppress the snake instability. Our analytical predictions are in good agreement with full numerical simulations of the Gross-Pitaevskii equation.

\end{abstract}

\maketitle



\section{Introduction}

Solitons, or solitary waves, are self-reinforcing nonlinear excitations that preserve their shape during propagation thanks to a delicate balance between dispersion and nonlinearity. They arise in a wide range of physical systems, including water waves, optical fibers, magnetic media, and atomic gases \cite{Korteweg01051895,Kivshar1998PR, Krkel1988, Weiner1988, Chen1993, FrantzeskakisJPA2010}. In self-defocusing (repulsive) nonlinear media, the soliton solution typically features a localized dip in density accompanied by a phase slip. These excitations are commonly referred to as dark solitons, particularly when the local density vanishes at the soliton core and the phase exhibits a sharp $\pi$ jump. The one-dimensional (1D) Gross-Pitaevskii (GP) equation, which governs the mean-field dynamics of Bose-Einstein condensates (BECs), supports such dark soliton solutions, and their experimental realization in ultracold atomic gases has opened a rich avenue for exploring nonlinear dynamics in quantum fluids \cite{Pethick-Smith, FrantzeskakisJPA2010}.

While solitons in quasi-1D geometries can exhibit remarkable stability, this robustness does not generally persist in higher dimensions. In two or three dimensions, dark solitons are susceptible to transverse modulations that cause their decay into more complex topological excitations. This decay process, known as the snake instability, is characterized by a bending of the soliton plane, eventually leading to the formation of vortex rings or solitonic vortices depending on the geometry and confinement of the system \cite{ZakharovJETP1974, Kivshar1998PR, Kivshar2000PR}. Early observations of this instability were made in nonlinear optical systems \cite{Mamaev1996PRL, Tikhonenko1996OL}, and subsequent experiments in BECs confirmed similar decay channels \cite{BurgerPRL1999, DenschlagSci2000, AndersonPRL2001, DuttonSci2001}. In particular, the nature of the decay products—vortex rings in isotropic traps or vortex lines in anisotropic geometries—has been shown to depend sensitively on the trap anisotropy, with theoretical predictions and experimental confirmations aligning well in this regard \cite{ Brand2002, KomineasPRA2003, DonadelloPRA2014, MithunJPCM2022, WellerPRL2008}.

\red{
The early theoretical investigations into the snake instability of dark soliton solutions in a homogeneous background \cite{KuznetsovJETP1988, Kuznetsov1995PRE} reveal the emergence of unstable modes with longer wavelengths beyond a critical length \cite{KuznetsovJETP1988}. These modes correspond to the exponential growth of transverse undulations leading to snake instability. Stabilization can be achieved by suppressing these long-wavelength unstable modes through the tuning of trap shapes, which has been extensively studied in various setups and geometries:
In a 3D setup, stabilization against transverse instability can be established by squeezing the condensate into a cigar shape, effectively creating a quasi-1D geometry in which dark solitons remain stable \cite{MuryshevPRA1999, FederPRA2000, KomineasPRA2003}. In 2D, reducing the transverse length of the dark soliton using infinite walls helps to suppress the unstable modes \cite{Brand2002}. However, most of these studies rely on the Bogoliubov-de Gennes (BdG) formalism, which}
requires linearization around a numerically obtained stationary solution, and does not easily lend itself to analytical treatment of dynamics or to capturing vortex formation in inhomogeneous systems. This motivates the search for alternative analytical frameworks \cite{kervekidis2004, Carretero_2008} capable of offering more intuitive and tractable descriptions of the instability dynamics \redsecond{\cite{Fetter-1, Fetter-2}}.

\red{
Building on the early insights, subsequent works have explored the instability and its decay dynamics in a variety of configurations, including cigar-shaped \cite{MithunJPCM2022}, disk-shaped \cite{Huang2003PRA, KevrekidisPRA2004, Verma2017PRA}, and hard-wall-confined condensates \cite{Hoefer2016PRA}. However, studies focusing specifically on 2D condensates with anisotropic confinement remain relatively scarce. Such setups are particularly relevant to experimental configurations, where the trap anisotropy can be directly controlled and quantified. This connection enables one to relate experimental quantities—such as the trap aspect ratio and central density—to the analytical framework. 

The connection between dark solitons and the vortex states that emerge following the instability has also been a subject of ongoing investigation. BdG analyses demonstrate that vortex dipole states bifurcate from the dark soliton branch in the excitation spectrum \cite{Stockhofe2011EPL}, indicating a crossover between these excitations \cite{Brand2001,Brand2002}. This connection has been further supported by recent dynamical studies, showing that dark solitons can transform adiabatically into vortex states \cite{Li2024PRA}, and that vortex oscillations exhibit soliton-like behavior beyond a critical anisotropy \cite{Tsatsos2016PRA}, providing additional evidence of their shared nonlinear origin. Moreover, long-lived oscillations between soliton states and their decay products near the onset of stabilization have been observed experimentally and predicted numerically \cite{KevrekidisPRA2004, Pietila2006PRA, Shomroni2009, Toikka2013PRA, Toikka2014JPB}. These oscillations are typically accompanied by the characteristic bending of the soliton, reminiscent of the snake instability. Altogether, these studies reveal a nontrivial connection between dark solitons and vortex states, in which nonlinearity plays a central role in governing their mutual dynamics and stability.

}

In this work, we present a variational approach \stfirst{to describe the snake instability of a dark soliton} \red{that simultaneously captures the snake instability and the crossover between dark solitons and vortex states} in an anisotropically trapped \red{2D} BEC. Our variational ansatz incorporates both the bending of the soliton plane and the onset of vortex-antivortex structures along the nodal line. This unified wavefunction \stfirst{allows us to analytically study} \red{provides an economical and fully analytical framework for investigating} the soliton's dynamical evolution \red{in non-uniform background densities} using the Lagrangian formalism. By deriving the Euler–Lagrange equations for the variational parameters, we identify \stfirst{both} stable oscillation modes, unstable directions in the parameter space\red{, and the full nonlinear dynamics of the system}. 

Our analysis yields critical values for trap anisotropy \red{\cite{Li2024PRA}}, beyond which the soliton remains dynamically stable. \stfirst{Our predictions are benchmarked against full numerical simulations of the time-dependent GP equation, and we find excellent agreement for both stable and unstable regimes.}\red{Stabilization occurs when the condensate exhibits an inverted parabolic profile, also known as the Thomas–Fermi (TF) profile. We also accurately reproduce nonlinear vortex–soliton oscillations, both of which match perfectly with full numerical simulations of the time-dependent GP equation. Importantly, it also allows the identification of unconditional soliton stability without fitting approximations\cite{KevrekidisPRA2004}, providing a reliable tool for analytical exploration of the instability.} 

\red{We analyze the system in three representative condensate density regimes: (i) a constant background density, which provides the simplest benchmark and allows us to identify the basic instability condition, (ii) the TF density profile, incorporating the effect of parabolic confinement and showing how the threshold depends on the varying density and the trap anisotropy, (iii) a Gaussian profile, describing the quasi-1D regime, where the transverse kinetic energy dominates and shows its impact on the stabilization of dark solitons against all the perturbation modes}.


The paper is organized as follows. Section~\ref{sec:model} describes the model system and the parameter regimes of interest. Section~\ref{sec:SI} presents numerical simulations showing the snake instability and introduces the variational wavefunction. Section~\ref{sec:dynamical}  analyzes the dynamics of the variational parameters through the Lagrangian approach. Section~\ref{sec:stability} focuses on the stability analysis and compares variational predictions with numerical results. The most unstable mode is investigated in Section~\ref{sec:most}. Finally, Section~\ref{sec:conc} summarizes our findings and outlines possible directions for future work.

\section{Model \label{sec:model}}

We consider $N$ bosons inside a harmonic potential 
\begin{equation}
\tilde V(\tilde{x},\tilde{y},\tilde{z}) = \frac{m}{2} (\tilde \omega_x^2 \tilde{x}^2 + \tilde \omega_y^2 \tilde{y}^2 + \tilde \omega_z^2 \tilde{z}^2),
\end{equation}
where $ m $ is the mass of a single atom and $ \tilde\omega_x, \tilde\omega_y, \tilde \omega_z $ are the trap frequencies in the corresponding directions. We assume strong confinement in the $ \tilde{z} $-direction, such that $ \tilde \omega_z \gg \tilde \omega_x, \tilde \omega_y $. Under this assumption, all particles occupy the ground state of the harmonic oscillator along the $ \tilde{z} $-axis. Integrating out the $ \tilde{z} $-dependence yields an effective two-dimensional description. For the remainder of this work, we study the dynamics of the two-dimensional condensate wavefunction $ \psi(\tilde{x},\tilde{y}) $, which is governed by the GP equation:
\begin{equation}\label{eqn:2D}
     -\frac{\hbar^2}{2 m} \tilde \nabla^2\psi + \tilde V(\boldsymbol{\tilde{r}}) \psi + g_{2D} |\psi|^2\psi = \mathrm i  \hbar \frac{\partial}{\partial \tilde{t}}\psi.
\end{equation}
The two dimensional interaction parameter is related to the s-wave contact interaction parameter $g_{3D}=\frac{4 \pi \hbar^2 a_s}{m} $ as  $g_{2D}\simeq g_{3D}/a_z $ where $a_s$ is the s-wave scattering length and $a_z$ is the oscillator length along the $\tilde z$-direction. 

We scale all lengths by the oscillator length in the x-direction $\ell=\sqrt{\frac{\hbar}{m \tilde{\omega}_x}}$, defining $x=\tilde{x}/\ell$, $y=\tilde{y}/\ell$; similarly all energies are scaled by $\hbar \tilde{\omega}_x$ and time is measured as $t=\tilde{\omega}_x \tilde{t}$. This allows us to express the time-independent GP equation as  
\begin{equation}\label{eqn:gpe}
-\frac{1}{2}\nabla^2\psi + \frac{1}{2} \left( x^2 + \omega^2 y^2 \right) \psi + g |\psi|^2\psi = \mu \psi.
\end{equation}
Here the trap anisotropy is $\omega=\tilde{\omega}_y/\tilde{\omega}_x$ and  the interaction strength is $g=\frac{m}{\hbar^2} g_{2D}$. The chemical potential $\mu$ is chosen so that $\int dx dy |\psi|^2 = N$.   From the scaled equation, one can see that the system is completely determined by the three dimensionless parameters $\omega, g$, and $N$. 

Typical cold atom condensates consist of a sufficiently large number of interacting atoms such that the condensate wavefunction varies slowly on the scale of the harmonic oscillator length. In this regime, the condensate density can be accurately described by the \stfirst{Thomas-Fermi (}TF\stfirst{)} approximation, where the kinetic energy contribution to the total energy is negligible compared to the interaction and potential energies. However, in an anisotropic trap, the applicability of the TF approximation depends on the degree of anisotropy, and two distinct limits emerge.

For small anisotropy, $ \omega \sim 1 $, where the trap is nearly isotropic in the transverse directions, the kinetic energy associated with wavefunction curvature can be neglected in both directions. This leads to a two-dimensional Thomas-Fermi (2DTF) density profile. In contrast, for large anisotropy $ \omega \gg 1 $, the condensate becomes quasi-one-dimensional. In this regime, the kinetic energy along the tightly confined direction is no longer negligible and shapes the density profile into a Gaussian, while the kinetic contribution along the weakly confined (long) direction remains small. This regime corresponds to the one-dimensional Thomas-Fermi (1DTF) limit. It is not a priori clear which regime—2DTF or 1DTF—is in play for stabilizing the soliton against snake instability. To address this, we analyze the system's behavior in both limits and determine the crossover anisotropy at which the dominant physical behavior transitions from one regime to the other.

In the 2DTF regime, the condensate wavefunction takes the form
\begin{equation}
    \psi(x,y)=\frac{1}{\sqrt{g}} \sqrt{\mu - \frac{x^2}{2} - \frac{\omega^2 y^2}{2}} \, \Theta\left( \mu - \frac{x^2}{2} - \frac{\omega^2 y^2}{2} \right),
\end{equation}
where $ \mu $ is the chemical potential and $ g $ is the effective two-dimensional interaction strength. This expression allows us to determine the chemical potential as $ \mu = \sqrt{\omega g N / \pi} $, and the TF radii as $ R_x = \sqrt{2\mu} $ and $ R_y = R_x / \omega $. The resulting central density in this regime is given by
\begin{equation}
   n_{2DTF}(0) = \sqrt{\frac{\omega N}{\pi g}}.
\end{equation}

In the 1DTF regime, the condensate is strongly confined along the $ y $-direction and exhibits a Gaussian profile with width $ \sigma_y = 1/\sqrt{\omega} $. In this limit, the transverse direction ($y$) is effectively frozen, and a one-dimensional GP equation describes the dynamics along the longitudinal axis. Solving the corresponding equation yields a central density with different scaling behavior:
\begin{equation}
   n_{1DTF}(0) = \left( \frac{9 \omega N^2}{16 \sqrt{2} \pi g} \right)^{1/3}.
\end{equation}

We solve the time-dependent GP equation~\eqref{eqn:gpe} using the split-step Fourier method \cite{Bao2003,Javanainen2006}. To obtain ground state configurations, we perform imaginary time evolution by implementing the Wick rotation $ t \rightarrow -\mathrm{i} \tau $. For all imaginary-time calculations, we initialize the condensate wavefunction with a uniform profile $ \psi_\text{i} = \sqrt{N / A} $, where $ A $ is the area of the computational domain.

\begin{figure}[t!]
    \centering
    \includegraphics[width=0.8\linewidth]{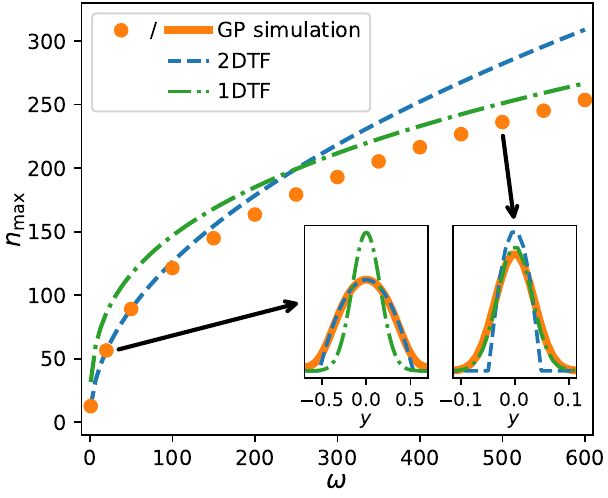}
    \caption{The maximum density $n_\text{max}$ as a function of the anisotropy ratio $\omega$ for \red{$gN=500$}. Numerically obtained GP simulations (dots) follow the 2DTF (dashed) approximation for smaller anisotropies and the 1DTF (dash-dotted) in the large anisotropy limit. The insets display cross-sections of the density profile along the $y$-axis for selected values of $\omega$, demonstrating the crossover from the 2DTF to the 1DTF regime in large anisotropies.}
    \label{fig:density_crossover}
\end{figure}

In Figure~\ref{fig:density_crossover}, we plot the central density obtained from numerical simulations as a function of the anisotropy parameter $ \omega $. The results clearly reveal a crossover from the 2DTF to the 1DTF regime. For small anisotropies, the numerics closely follow the 2DTF scaling, whereas for large anisotropies, the behavior transitions to match the 1DTF prediction. \red{One expects a crossover from 2D to a quasi-1D regime at densities corresponding to $\mu=g n_0\sim\omega$ \cite{Petrov2004}, which results with the expectation of dimensional crossover anisotropy around $\omega_d\sim0.318\ gN$. Indeed,} the crossover occurs at
\begin{equation} \label{eqn:omega_s}
\red{\omega_d} = \pi \left( \frac{9}{16 \sqrt{2}} \right)^2 g N \simeq 0.497\, g N,
\end{equation}
which corresponds to the intersection point of the two analytical expressions. Additionally, we verify that the full density profiles obtained numerically closely follow the TF forms deep within both limiting regimes.

\red{Here, we emphasize that the dimensional crossover anisotropy \red{$\omega_d$} given in Eq.~\eqref{eqn:omega_s} describes only the point where the density profile of the
ground-state changes. In the later sections, we will use the value of \red{$\omega_d$} to determine what shape we expect for the ground state of the condensate for a given $gN$ inside a harmonic trap with a given anisotropy $\omega$.}

\section{Variational Wavefunction for Snake Instability \label{sec:SI}}

\begin{figure}[h!]
    \centering
    \includegraphics[width=0.99\linewidth]{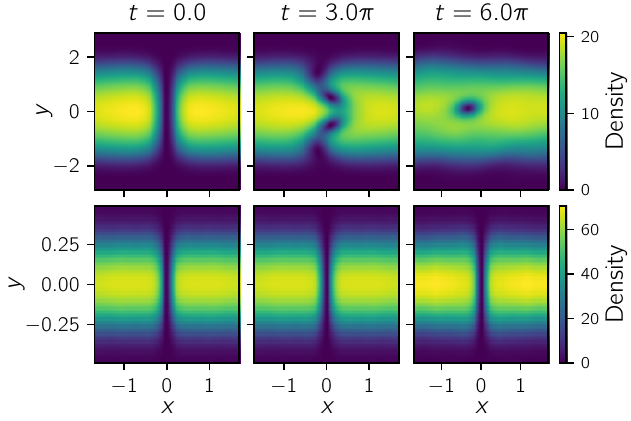}
    \caption{Time evolution of a dark soliton in a trapped Bose-Einstein condensate for two different trap anisotropies, both with $ g = 1 $ and $ N = 400 $. The colorbars show the density of BECs. Top row: For $ \omega = 3 $, the soliton is dynamically unstable and decays via the snake instability, leading to the formation of vortex structures. Bottom row: For $ \omega = 35 $, the soliton remains stable and persists without significant deformation. Note that the vertical axis ($ y $) is scaled differently in the two panels to reflect the change in trap geometry.}
    \label{fig:SI}
\end{figure}

To obtain a stationary soliton configuration within the trapped condensate, we initialize the wavefunction with a nodal line along the $ x=0 $ axis and enforce odd parity under reflection $ x \to -x $. This symmetry condition ensures the presence of a dark soliton centered at $ x=0 $, as the density must vanish along this plane. We then evolve the GP equation in imaginary time. This relaxes the system to a metastable state consistent with the imposed symmetry. The resulting wavefunction corresponds to a dark soliton embedded within the anisotropic ground-state density profile of the condensate.

The soliton profile in the absence of external trapping is well understood in the one-dimensional GP equation,
\begin{equation}
    - \frac{1}{2} \frac{d^2\psi}{dx^2} + g |\psi|^2 \psi = \mu \psi,
\end{equation}
which admits the exact solution $ \psi(x) = A \tanh(x/\xi) $, with amplitude $ A = \sqrt{\mu/g} $ and healing length $ \xi = 1/\sqrt{\mu} $. Our numerical soliton solutions in the trapped two-dimensional geometry exhibit a similar structure, with the wavefunction closely approximated by
\begin{equation}
    \psi(x,y) \approx \sqrt{n(x,y)} \tanh(x/\xi),
\end{equation}
where $ n(x,y) $ represents a slowly varying envelope consistent with the background condensate density. This form confirms that the nodal structure of the soliton is retained even in the presence of trapping and anisotropy.

To assess the stability of the soliton, we evolve the system in real time starting from the stationary soliton configuration. The presence of small perturbations—whether from numerical noise or residual excitation—provides a natural seed for instability. If the soliton is unstable, we observe the nodal plane bending over time and eventually breaking into more complex structures such as vortices. In contrast, a stable soliton maintains its profile with negligible deformation for extended durations. We find that simulating the dynamics over a time window corresponding to four to six trap periods in the $ x $-direction ($ t \sim 4\pi $ to $ 6\pi $) is sufficient to determine stability. Figure~\ref{fig:SI} illustrates this behavior: the top panel shows the soliton bending and decaying into vortex structures for a weakly anisotropic trap ($ \omega = 2 $), while the bottom panel displays a long-lived, stable soliton for a strongly anisotropic trap ($ \omega = 35 $). Note that the vertical ($ y $) axis is scaled differently in the two panels to account for the varying trap anisotropy.

The snake instability manifests itself through two intertwined mechanisms: the transverse bending of the soliton plane and the subsequent fragmentation of the nodal line into point-like topological defects. The soliton, initially characterized by a planar density node and a localized phase jump, becomes dynamically unstable when transverse modulations are energetically favorable. As the soliton bends, regions of higher curvature incur a kinetic energy cost due to increased phase gradients. In a condensate that is sufficiently narrow along the transverse direction, there is simply not enough spatial extent for such bending to occur without excessive energy penalty, leading to dynamical stabilization of the soliton.

However, the bending of the nodal plane is not the sole feature of the instability. The nodal line itself is susceptible to fragment into a sequence of point nodes, each accompanied by localized phase winding. These point defects correspond to vortex-antivortex pairs aligned along the former soliton plane. Viewed from a distance, the alternating vortices collectively \stsecond{maintain the overall phase structure imposed by}\redsecond{give rise to a phase pattern that is qualitatively consistent with that of} the original soliton, including the characteristic $\pi$ phase shift across the nodal region \red{\cite{Verma2017PRA, Shomroni2009, Hoefer2016PRA}}. The vortex cores may appear anisotropic due to the geometry of the trap and the inhomogeneity of the condensate. Our numerical simulations reveal both bending and vortex formation, in agreement with experimental observations of dark soliton decay in trapped BECs \cite{DuttonSci2001}.

\begin{figure}[!h]
    \centering
    \includegraphics[width=0.95\linewidth]{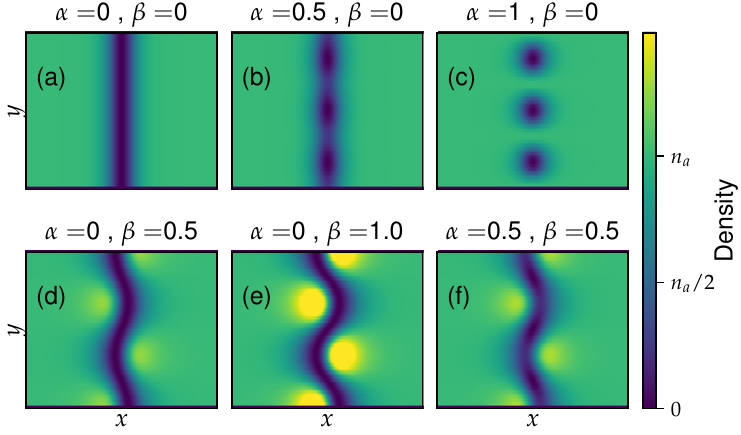}
    \caption{Density profiles of variational wavefunction defined in Eq.~\ref{eqn:ansatz-crossover} with constant density $n_a$ for various choices of $\alpha$ and $\beta$. The colorbar shows the density of the wavefunction. The parameter $\alpha$ controls the nodal line breaking up into vortices, while $\beta$ determines the bending amplitude of the soliton plane.}
    \label{fig:ansatz-densities}
\end{figure}

\begin{figure}[!h]
    \centering
    \includegraphics[width=0.8\linewidth]{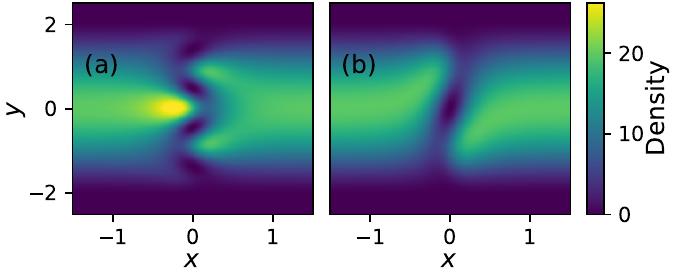}
    \caption{\red{Density profiles of variational wavefunction defined in Eq.~\ref{eqn:ansatz-crossover} inside a TF-profile background density for $\alpha=0.7$ and $\beta=0.4$. The wavevector $k$ controls the spacing between vortices, which is related to the number of vortices inside the condensate. We show two modes corresponding to (a) four vortices and (b) a single vortex as a decay product of snake instability inside an anisotropic condensate with $gN=400$ and $\omega=3$, as used in the first row of Fig.~\ref{fig:SI}.}}
    \label{fig:ansatz-densities-envelope}
\end{figure}

To analytically capture the essential features of the snake instability, we construct a variational ansatz that incorporates both bending of the soliton plane and the emergence of vortex-like excitations along the nodal line. Our proposed wavefunction is
\begin{equation} \label{eqn:ansatz-crossover}
\psi(x,y) = \sqrt{n(x,y)} \bigg( \tanh\left(\frac{x}{\xi_s} \right) + \mathrm i A\,  \frac{\sin(ky)}{\mathrm{cosh} \left(x/\xi_s \right)} \bigg),
\end{equation}
where $ n(x,y) $ denotes the background condensate density, $ \xi_s $ is the soliton coherence length, and $ k $ is the wavenumber characterizing modulations along the soliton plane. The variational parameter $ A = \alpha + i\beta $ governs the deviation from the stationary soliton. When $ A = 0 $, the ansatz reduces to a pure dark soliton with a planar nodal line. A nonzero imaginary part $ \beta $ introduces a bending of the soliton plane, corresponding to the snake-like undulation observed in simulations. Meanwhile, the real part $ \alpha $ breaks the nodal line into a chain of alternating phase singularities—effectively a necklace of vortices and antivortices. These features qualitatively reproduce the key behaviors seen in both numerical simulations and experimental observations. Figure~\ref{fig:ansatz-densities} illustrates how varying the parameter $ A $ leads to different density configurations, ranging from a straight soliton to vortex-laden states. \red{In a similar sense, Figure~\ref{fig:ansatz-densities-envelope} illustrates how varying the wavevector of the mode $k$ controls the spacing between successive vortices. While the wavelength of the mode $\lambda=\frac{2\pi}{k}$ controls the separation of the appearing vortices, the ratio between the size of a squeezed condensate, as in Figure~\ref{fig:ansatz-densities-envelope}, and the wavelength $\lambda$ determines the number of vortices appearing on the non-uniform background density.}

In the following analysis, we treat $ A $ as a dynamical variable within a Lagrangian framework and use it to characterize the onset and growth of the snake instability.  As long as $ |A| $ remains small, we qualitatively capture both the bending of the soliton plane and the breakup into alternating vortices within a unified description. The $\mathrm{sech}(x/\xi_s)$ factor in the ansatz ensures that the perturbations remain localized around the original soliton plane, which not only reflects the physical nature of the instability but also simplifies the analysis by allowing integrals over powers of the wavefunction to be evaluated in closed form using elementary functions.

To analyze the dynamics of the instability, we adopt the time-dependent variational principle based on a Lagrangian formulation. Substituting the ansatz~\eqref{eqn:ansatz-crossover} into the Lagrangian density corresponding to the GP equation and integrating over the spatial coordinates yields an effective Lagrangian $ L(\alpha, \beta, \dot{\alpha}, \dot{\beta}) $ for the variational parameters. We have $ L = T - U$ with: 
\begin{subequations} \label{eqn:L}
    \begin{gather}
        T = \frac{\mathrm i}{2}\int \mathrm dx \mathrm dy \left( \psi^* \dot \psi - \psi \dot \psi^* \right)  \label{eqn:T}, \\
        U = \int \mathrm dx \mathrm dy\left(\frac{|\nabla \psi|^2}{2} + V(\boldsymbol r)|\psi|^2 + \frac{g|\psi|^4}{2}-\mu |\psi|^2\right).  \label{eqn:U}
    \end{gather}
\end{subequations}

In the next section, we evaluate the kinetic, potential, and interaction energy contributions to the Lagrangian separately and present the resulting equations governing the growth and oscillation of the perturbation amplitude $ A(t) $. We then derive the equations of motion for $ \alpha(t) $ and $ \beta(t) $ using the Euler–Lagrange equations. This approach provides a low-dimensional dynamical system that captures the onset of the instability as a function of the trap anisotropy and the modulation wavenumber $ k $.

\section{\label{sec:dynamical}Dynamical Equations}

The effective Lagrangian derived from our variational ansatz depends on the background condensate density $ n(x,y) $, which differs significantly between the 1DTF and 2DTF regimes. In particular, the spatial profile of $ n(x,y) $ influences the weighting of kinetic, potential, and interaction energy contributions, thereby modifying the dynamics of the perturbation amplitude $ A(t) $. To isolate the essential features of the instability mechanism and identify the relevant regime for soliton stabilization, we begin by adopting the simplest approximation: a uniform background density $ n(x,y) = n_a $.

This constant-density assumption enables closed-form evaluation of all spatial integrals in the Lagrangian, providing a transparent analytical expression for the effective dynamics. Although approximate, this approach captures the core physics of the snake instability and allows us to determine whether stabilization arises within the quasi-one-dimensional nature of the 1DTF regime or the broader spatial extent of the 2DTF profile.
We introduce a length $\ell_y$ along the $y$ direction to limit our integration region, and substituting the ansatz~\eqref{eqn:ansatz-crossover} with $ n(x,y) = n_a $ into the kinetic, potential, and interaction terms of the Lagrangian in Eqs.~\eqref{eqn:L}, we obtain the following expressions for each energy contribution:
\begin{subequations} \label{eqn:lagrangian-terms}
\begin{align}
T &= n_a \xi_s \ell_y \frac{\mathrm i}{2}\left(A^* \dot{A} - A \dot{A^*}\right),  \\
U_\text{kin} &= - \frac{k^2}{2}  n_a \xi_s \ell_y |A|^2 - \frac{1}{6 \xi_s^2} n_a \xi_s \ell_y |A|^2,  \\
U_\mu &= + \mu n_a \xi_s \ell_y |A|^2,  \\
U_\text{int} &=  - \frac{1}{4} g n_a^2 \xi_s \ell_y |A|^4 - \frac{1}{3} g n_a^2 \xi_s \ell_y |A|^2 - \frac{2}{3} g n_a^2 \xi_s \ell_y \mathrm{Im}(A)^2. 
\end{align}
\end{subequations}

We treat the real and imaginary parts of the variational parameter $ A = \alpha + i\beta $ as generalized coordinates and derive the equations of motion using the Euler–Lagrange formalism. This yields two coupled differential equations governing the time evolution of $ \alpha(t) $ and $ \beta(t) $. To simplify the resulting expressions, we express all quantities in terms of the chemical potential $ \mu $, using $ n_a = \mu / g $ for the background density and $ \xi_s^{-2} = \mu $ for the soliton coherence length. After substitution and simplification, we obtain the following equations of motion:
\begin{subequations}  \label{eqn:const_density-eom}
\begin{align}
2 \dot{\alpha} &= -\left( \mu - k^2 -\frac{4}{3} \mu \right) \beta +  \mu \left(  \beta^3 +  \alpha^2 \beta \right),   \\
2 \dot{\beta} &= \left( \mu - k^2 \right) \alpha - \mu \left(  \alpha^3 +  \beta^2 \alpha \right).
\end{align}
\end{subequations}

\begin{figure}[h!]
    \centering
    \includegraphics[width=0.95\linewidth]{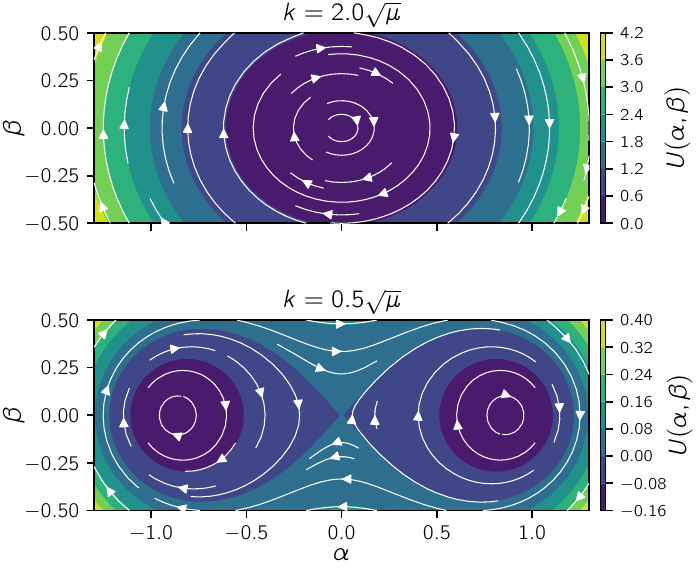}
    \caption{Phase space of the variational dynamical system defined by Eqs.\eqref{eqn:const_density-eom} in the complex $A=\alpha+\mathrm i\beta$ plane, illustrating the dependence of the flow topology on the perturbation wavenumber $k$. The colorbars indicate the potential energy $U(\alpha,\beta)$ calculated with Eq.~\eqref{eqn:U}, and the arrows show the direction of the flow in the phase space. For large $k=2.0\sqrt{\mu}$,(top), the origin is a stable center, and all nearby trajectories form closed orbits, indicating dynamically stable soliton configurations. For small $k=0.5\sqrt{\mu}$(bottom), the origin becomes a saddle point and two stable fixed points appear along the $\beta=0$ axis, signaling stable necklace of vortices and antivortices.}
    \label{fig:phase_space_plot}
\end{figure}

The dynamical system described by Eqs.~\eqref{eqn:const_density-eom} defines a two-dimensional flow in the phase space spanned by the real and imaginary parts of the variational parameter $ A = \alpha + i\beta $. The topology of this phase space depends on the value of the perturbation wavenumber $ k $ along the soliton plane. As shown in Fig.~\ref{fig:phase_space_plot}, there are two qualitatively distinct regimes. For large values of $ k $, the origin $ (\alpha,\beta) = (0,0) $ is the only fixed point, and all nearby trajectories form closed orbits encircling it. This indicates stable, bounded oscillations of the variational parameter $ A(t) $, and corresponds to a soliton that resists both bending and vortex formation. In this regime, the soliton is dynamically stable against the snake instability.

In contrast, for small $ k $, the phase space undergoes a bifurcation: the origin becomes a saddle point, and two additional stable fixed points emerge along the $ \beta = 0 $ line at finite $ \alpha $. These off-center fixed points correspond to stationary configurations where the nodal line of the soliton has broken up into a regular array of vortex-antivortex pairs—a structure we refer to as a vortex-antivortex necklace. In this regime, trajectories initialized near the origin are deflected away and encircle the off-axis fixed points instead. Along these trajectories, the magnitude $ |A| $ grows significantly, violating the small-amplitude assumption used in our variational analysis. We therefore interpret the transition in phase space topology—from a single center to a three-fixed-point structure—as the signature of the snake instability. The critical value of $ k $ at which this bifurcation occurs defines the onset of instability within our variational model.

To examine the stability of the dark soliton solution, we linearize the equations of motion around the stationary point $ A = 0 $, assuming small perturbations. Retaining only leading-order terms in $ \beta $, we obtain the second-order differential equation
\begin{equation}
    4 \ddot{\beta} \approx -\left(\mu - k^2\right)\left(-\frac{1}{3}\mu - k^2\right) \beta.
\end{equation}
This expression reveals the stability criterion of the soliton with respect to transverse modulations. For wavevectors satisfying $ k^2 > \mu $, the coefficient is negative and solutions correspond to bounded oscillations—indicating dynamical stability. Conversely, for $ k^2 < \mu $, the sign of the coefficient becomes positive over part of the range, and exponentially growing solutions emerge \cite{KuznetsovJETP1988}. The maximum growth rate occurs at $ k = \sqrt{\mu/3} $, which defines the most unstable mode. For stable oscillations, the frequency is given by
\begin{equation}
\label{eqn:constant_oscillation}
    \Omega =\frac{1}{2}\sqrt{\left(\mu - k^2\right)\left(-\frac{1}{3}\mu - k^2\right)}.
\end{equation}

\begin{figure}[h!]
   \centering
   \includegraphics[width=0.95\linewidth]{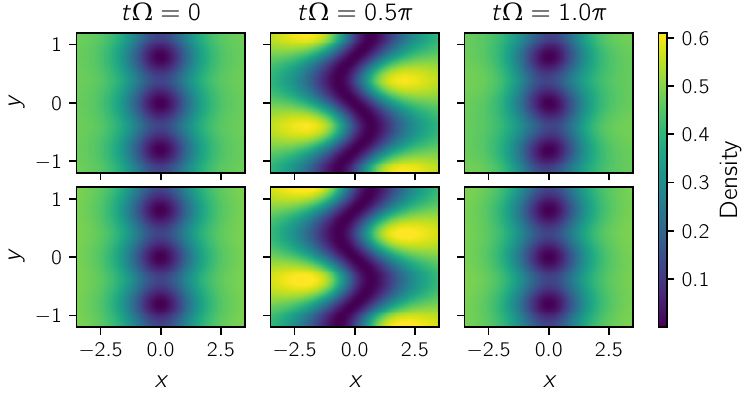}
   \caption{Stable oscillation of a dark soliton mode with finite modulation $ k = 4 $. The colorbar shows the density.
   We set the chemical potential $\mu=0.5$ and the initial value for the parameter $A$ as $0.5$. 
   The time evolution of the GP state (top row) is accurately reproduced by our variational ansatz wavefunction (bottom row), which is evolved according to Eqs.\eqref{eqn:const_density-eom}.
   The observed oscillation period agrees well with the analytical prediction from Eq.~\eqref{eqn:constant_oscillation}, showing a recurrence of the initial necklace of vortices and antivortices at each half period $t\Omega=\pi$ around the origin in the parameter space. We verified that the GP state and our ansatz wavefunction remain in close agreement up to at least $t\Omega=7\pi$.}
   \label{fig:oscillation}
\end{figure}

\begin{figure}[h!]
   \centering
   \includegraphics[width=0.98\linewidth]{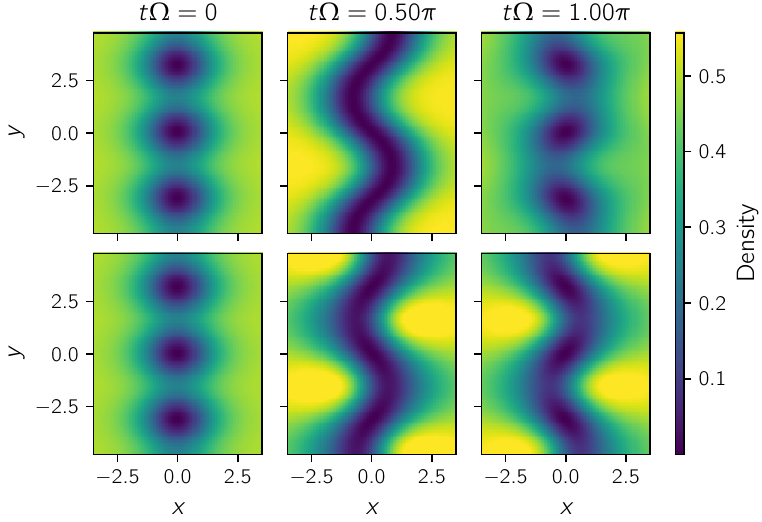}
   \caption{\red{Similar to Fig.~6, but with stronger nonlinear effects. The parameters are, modulation $k = 1$, chemical potential $\mu = 0.5$, and initial value $A = 0.7$. The parameters are selected such that the ratio between the nonlinear contributions in Eqs.~\eqref{eqn:const_density-eom} to the linear ones is as high as $1/2$ during the oscillation, which shows that our ansatz wavefunction in Eq.~\eqref{eqn:ansatz-crossover} agrees well with the GP simulations, although quantitative agreement begins to deteriorate at \stsecond{shorther}\redsecond{longer} time scales.
   }}
   \label{fig:oscillation-nonlinear}
\end{figure}

To verify the accuracy of our variational model, we compare its predictions with numerical simulations of the GP equation. In the stable regime, we simulate a soliton embedded in a condensate subject to periodic boundary conditions along the $ y $-direction and monitor the temporal evolution of its nodal structure. As illustrated in Fig.~\ref{fig:oscillation}, we observe coherent oscillations of the soliton plane. The period extracted from the simulation matches closely with the analytical prediction from the linearized theory, confirming the validity of the variational approach for small deviations from the stationary soliton configuration. 
\red{
Similarly, Fig.~\ref{fig:oscillation-nonlinear} illustrates the oscillation of the mode with $k$ closer to the critical value $k_c=\sqrt{\mu}$, where nonlinear effects strongly influence the dynamics governed by Eqs.~\eqref{eqn:const_density-eom}. The oscillation frequency $\Omega$ is noticeably shifted due to the impact of nonlinearity, 
leading to the formation of an oriented vortex–antivortex necklace at $t\Omega=\pi$ in Fig.~\ref{fig:oscillation-nonlinear}, rather than a straight alignment. 
Although the frequency obtained from our variational equations deviates quantitatively from the full GP result, the model successfully captures the stable soliton–vortex oscillations even beyond the linear regime ($A \rightarrow 0$ or $k\gg\sqrt{\mu}$). The discrepancies arise from the reduction of $\Omega$ as $k \rightarrow k_c = \sqrt{\mu}$, resulting in longer oscillation periods during which additional excitations, such as sound waves, can emerge in the condensate.
}

We now use the analytical results derived from the linearized dynamics to estimate the critical trap anisotropy required for stabilizing the dark soliton. The key assumption is that unstable transverse modes can only be excited if their wavelength fits within the transverse extent of the condensate. In particular, we assume that modes with wavelength larger than the size of the condensate along the $ y $-direction are suppressed \cite{KevrekidisPRA2004}. Using this physical intuition, we estimate the stabilization threshold by requiring that the smallest wavelength unstable mode—characterized by wavevector $ k = \sqrt{\mu} $ and corresponding wavelength $ \lambda_{\text{max}} = 2\pi/k $—no longer fits within the condensate. In the TF limit, the transverse extent of the condensate is given by $ 2R_y = 2 \sqrt{2\mu}/\omega $, and demanding $ 2R_y \lesssim \lambda_{\text{max}} $ yields a condition on the critical anisotropy.

To make this estimate explicit, we adopt the 2DTF approximation and substitute the analytical expression for the chemical potential, $ \mu = \sqrt{\omega g N / \pi} $, obtained in Eq.~(9). Solving $ 2R_y \simeq \lambda_{\text{max}} = 2\pi/\sqrt{\mu} $ then gives the critical \red{stabilization} anisotropy
\begin{equation}\label{eqn:omega_c-constant_density}
    \omega_c = \frac{2}{\pi^3 }g N \approx 0.0645\, g N.
\end{equation}
This result indicates that the soliton is stabilized against the snake instability for anisotropies larger than $ \omega_c $. Notably, $ \omega_c $ is smaller than \red{the dimensional crossover anisotropy} $ \red{\omega_d} \approx 0.497\, g N $ identified earlier, where the system transitions from the 2DTF to the 1DTF regime. This suggests that stabilization occurs \stfirst{well before the condensate enters the quasi-one-dimensional limit} \red{in the deep 2DTF regime}, consistent with our observation that the soliton becomes dynamically stable while the condensate density is still well described by the 2DTF profile.
In the next section, we refine our variational approach by incorporating the spatial inhomogeneity of the condensate density, assuming a 2DTF background profile for $ n(x,y) $.

\section{Stability of the soliton \label{sec:stability}}

To improve upon the constant-density approximation used in the previous section, we now incorporate the spatial inhomogeneity of the condensate by using the TF density profile as the background for our variational ansatz. This refinement allows us to more accurately capture how the condensate's confinement affects the energetics and the stability of the soliton. In particular, we examine how the effective Lagrangian depends on the trap anisotropy through the position-dependent density $ n(x,y) $ characteristic of the 2DTF regime.

We now substitute the variational ansatz
\begin{equation} \label{eqn:ansatz-TF}
\psi(x,y,t) = \sqrt{n} \sqrt{1 - \frac{y^2}{R_y^2}} \left( \tanh\left(\frac{x}{\xi_s}\right) + i A(t) \frac{\sin(k y)}{\cosh\left(x/\xi_s\right)} \right),
\end{equation}
into the Lagrangian formulation of the GP equation. Here $ n $ is a normalization constant, and $ R_y = \sqrt{2\mu}/\omega $ is the transverse TF radius of the condensate. This ansatz captures the spatial inhomogeneity of the condensate density in the transverse ($ y $) direction while maintaining analytical tractability in the $ x $-direction through the localized solitonic structure. Unlike in the constant-density case, the external harmonic potential in the $ y $-direction now contributes explicitly to the potential energy term in the Lagrangian.

To proceed analytically, we adopt the standard TF approximation, which assumes that the background density profile varies slowly compared to the solitonic perturbation. Consequently, when evaluating derivatives with respect to $ y $, we act only on the perturbation term $ \sin(k y) $, not on the background density $ \sqrt{1 - y^2/R_y^2} $. 

Evaluating the effective Lagrangian with the TF background and the variational ansatz introduced above, we obtain a closed-form expression for $ L $ in terms of the variational parameter $ A(t) = \alpha(t) + i \beta(t) $. The integrals over the $ x $-direction are computed analytically using standard identities for hyperbolic functions, while the $ y $-dependence is retained through the background envelope $ \sqrt{1 - y^2/R_y^2} $. We neglect derivatives of the background density in accordance with the TF approximation \red{and the confinement along the $x$-direction because of its negligible contribution}. The resulting Lagrangian after reducing a common factor of $n \xi_s R_y$ reads:
\begin{align}
L ={}& F(k R_y)\,  (\beta \dot{\alpha} - \alpha \dot{\beta}) \nonumber \\
&+ F(k R_y)\, \left(\mu-\frac{k^2}{2} - \frac{\xi_s^{-2}}{6} \right) (\alpha^2+\beta^2) \nonumber \\
& - \omega^2 R_y^2 G_1(k R_y) \, (\alpha^2+ \beta^2) \\
&- \frac{2}{3} g n G_3(k R_y) (3 \beta^2+\alpha^2)  \nonumber \\
&- \frac{2}{3}g n G_2(k R_y) \, (\alpha^2+\beta^2)^2, \nonumber
\end{align}
where the functions $ F, G_1,G_2,G_3 $ depend on the dimensionless combination of the transverse TF radius $ R_y $ with wavevector $k$. These functions  are:
\begin{align}
F(x) &= \frac{4}{3} +\frac{1}{x^2} \cos(2x)-\frac{1}{2x^3}\sin(2x), \label{eq:F} \\
G_1(x) &= \frac{2}{15} +\frac{x^2-3}{2x^4} \cos(2x)+\frac{3-5x^2}{4x^5}\sin(2x),  \label{eq:G1} \\
G_2(x) &=\frac{2}{5}  + \frac{(3-16x^2) \sin (4 x)}{512 x^5}-\frac{(3-4x^2) \sin (2 x)}{4 x^5} \nonumber \\
&+\frac{3 \cos (2 x)}{2 x^4}-\frac{3 \cos (4 x)}{128 x^4}, \label{eq:G2} \\
G_3(x) &= \frac{8}{15} +\frac{3}{2x^4} \cos(2x)+\frac{4x^2-3}{4x^5}\sin(2x).  \label{eq:G3}
\end{align}
The details of the calculations are given in the appendix. The oscillatory character of these functions shows that the Lagrangian has some control over how the perturbation fits into the transverse direction of the condensate. Noticing $G_1(x)+G_3(x)=F(x)/2$ simplifies the derivation of the dynamical equations.

The inclusion of the TF background modifies the effective coefficients in the Lagrangian, but it does not qualitatively alter the structure of the dynamical phase space. As in the constant-density case, the dynamics of the variational parameters $ \alpha(t) $ and $ \beta(t) $ are governed by a two-dimensional nonlinear system derived from the Euler–Lagrange equations. While the specific trajectories and growth rates are now modulated by the functions $ F(k R_y) $, $ G_1(k R_y) $,$ G_2(k R_y) $, and $ G_3(k R_y) $, the overall topology of the flow—featuring a transition from bounded oscillations to instability—remains intact. In particular, the origin $ (\alpha, \beta) = (0,0) $, corresponding to the unperturbed planar soliton, remains a fixed point of the dynamics.

To assess the stability of this fixed point, we linearize the equations of motion around the origin. Retaining only terms up to second order in $ \alpha $ and $ \beta $, we find that their time evolution is governed by:
\begin{align}
\dot{\alpha} &= -  \left[\mu - \frac{k^2}{2} - \frac{ \xi_s^{-2}}{6} -\frac{\omega^2 R_y^2 G_1(k R_y)}{F(k R_y)}-\frac{2 g n G_3(k R_y)}{ F(k R_y)}\right]  \beta, \label{eq:TF_lin_alpha} \\
\dot{\beta}  &= + \left[\mu - \frac{k^2}{2} - \frac{ \xi_s^{-2}}{6} -\frac{\omega^2 R_y^2 G_1(k R_y)}{F(k R_y)}-\frac{2 g n G_3(k R_y)}{3 F(k R_y)}\right] \alpha. \label{eq:TF_lin_beta}
\end{align}

To simplify the stability criterion and reveal its dependence on a single physical scale, we now substitute the 2DTF identities $ \xi_s^{-2} = \mu $, $ R_y^2 \omega^2 = 2\mu $, and $ g n = \mu $ into Eqs.~\eqref{eq:TF_lin_alpha}–\eqref{eq:TF_lin_beta}.  With these substitutions, the linearized equations reduce to a compact form where the oscillation frequency depends only on the dimensionless combination $ k R_y $ through $G_1(k R_y),G_3(k R_y)$ and $F(k R_y)$:
\begin{equation}
\Omega^2(k) = \mu \left( -\frac{1}{6} - \frac{k^2}{2 \mu} \right) \left( \frac{5}{6} - \frac{k^2}{2 \mu}- 2 \frac{G_1+G_3/3}{F}  \right) . \label{eq:TF_freq}
\end{equation}
 The soliton is dynamically stable for values of $ k $ such that $ \Omega^2(k) > 0 $, while modes with $ \Omega^2(k) < 0 $ are unstable and grow exponentially in time. 

Although the Lagrangian formulation incorporates the effect of transverse confinement through the functions $ F(k R_y) $, $ G_1(k R_y) $, and $ G_3(k R_y) $, it does not yield unconditional stability across all wavenumbers. In particular, low-$ k $ modes remain unstable for any choice of parameters, as the oscillation frequency \eqref{eq:TF_freq} becomes imaginary in this regime. Consequently, to define a stability threshold, $\omega_c$, we must select a representative value of $ k R_y $ at which to evaluate the onset of instability. 

To make this choice physically motivated, we consider the fact that in the TF regime, the local chemical potential varies across the transverse direction as $ \mu(y) = \mu \left(1 - y^2/R_y^2\right) $. In the constant-density approximation, the last unstable mode corresponds to $ k \simeq \sqrt{\mu} $. In analogy, we now take the transverse average of the local healing scale and define an effective wavenumber
\begin{equation}
k = \frac{\sqrt{\mu}}{2R_y} \int_{-R_y}^{R_y} \sqrt{1 - \frac{y^2}{R_y^2}} \, dy = \sqrt{\mu} \cdot \frac{\pi}{4},
\end{equation}
which yields $ k R_y = 2 $ in dimensionless units. Substituting this value into Eq.~\eqref{eq:TF_freq} and solving $ \Omega^2(k) = 0 $ defines the critical \red{stabilization} anisotropy:
\begin{equation} \label{eqn:omega_c-TF}
\omega_c = \left(\frac{1}{2} - \frac{4 G_1(2)}{3 F(2)}\right) g N \simeq 0.0847\, g N,
\end{equation}
\red{which corresponds to the relation between harmonic length $a_y$ and the condensate healing length $\xi=1/\sqrt{2\mu}$ as $a_y\simeq1.969\,\xi$, as reported in a recent study \cite{Li2024PRA}.}
\stfirst{This estimate represents an improved stabilization threshold that}
\red{Unlike the estimate based on a uniform density in Eq.~\eqref{eqn:omega_c-constant_density}, the critical \red{stabilization} anisotropy in Eq.~\eqref{eqn:omega_c-TF}} incorporates both the inhomogeneous background density and the spatial structure of the perturbation, \red{which are the effects relevant to experiments.}

To validate the analytical predictions and accurately determine the critical \red{stabilization} anisotropy\stfirst{for stabilization}, we perform time-dependent GP simulations for varying trap geometries. For each value of anisotropy $ \omega $, we begin by preparing a stationary soliton configuration\red{, as described in Sec.~III.} \stfirst{This is achieved by imposing odd parity under reflection $ x \rightarrow -x $, which enforces a nodal plane at $ x = 0 $, and evolving the GP equation in imaginary time. This procedure yields a metastable soliton state consistent with the imposed symmetry and the given trap parameters.}
Once the initial soliton state is prepared, we evolve the system in real time and monitor its dynamical stability. To quantify the persistence of the soliton, we define a time-dependent overlap function
\begin{equation} \label{eqn:Ft}
F(t) = \left| \int\mathrm dx\,\mathrm dy\, \psi^*(x,y,t)\, \psi(x,y,0)\, \frac{1}{\cosh^2(x/\xi_s)} \right|,
\end{equation}
which measures the projection of the evolving wavefunction onto the initial soliton state, weighted by the localized profile of the soliton. For a stable soliton, this quantity exhibits bounded oscillations over time, reflecting coherent motion of the perturbation. In contrast, for an unstable soliton, $ F(t) $ remains initially steady but then exhibits an abrupt decay at a time determined by the growth of unstable modes seeded by numerical noise.

\begin{figure}[!t]
    \centering
    \includegraphics[width=0.95\linewidth]{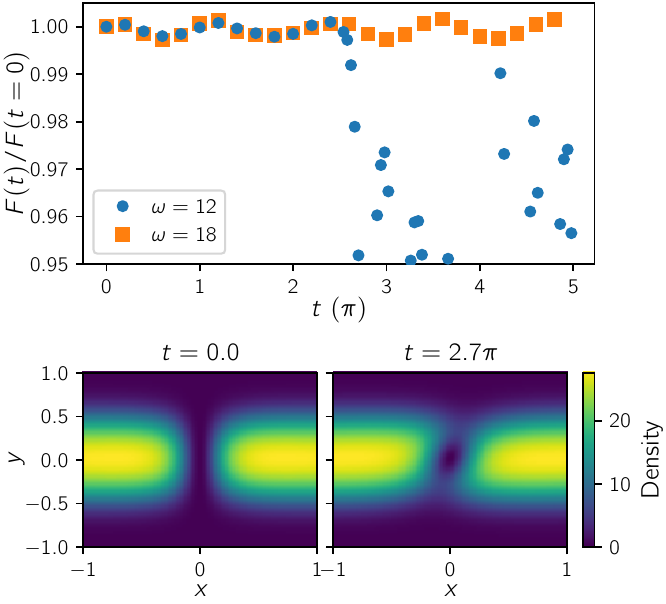}
    \caption{(Top) The time evolution of the overlap function $F(t)$ defined in Eq.~\eqref{eqn:Ft} for two different trap anisotropies and $gN=200$. Below the critical \red{stabilization} anisotropy (circle), the dark soliton decays into vortices and $F(t)$ rapidly drops below 1\% of its initial value, whereas beyond the critical value (square), the overlap $F$ stays almost the same. (Bottom row) The dynamical evolution of the unstable configuration in the first figure ($\omega=12$). The colorbar shows the density. Approaching the critical \red{stabilization} anisotropy, the unstable dark soliton decays through an odd-in-y perturbation with a single vortex at the center.}
    \label{fig:Ft_plot}
\end{figure}

Figure~\ref{fig:Ft_plot} illustrates this behavior by showing the time evolution of $ F(t) $ for two representative cases: one in the stable regime and one in the unstable regime. We define the soliton to be dynamically stable if the value of $ F(t) $ remains within 1\% of its initial value over a duration of $ \omega_x t = 5\pi $, corresponding to several trap periods. While the exact threshold for stability could in principle depend on the length of the simulation and the sensitivity to numerical noise, we find that this criterion is robust given our numerical accuracy and observed error accumulation.

\begin{figure}[h!]
    \centering
    \includegraphics[width=0.9\linewidth]{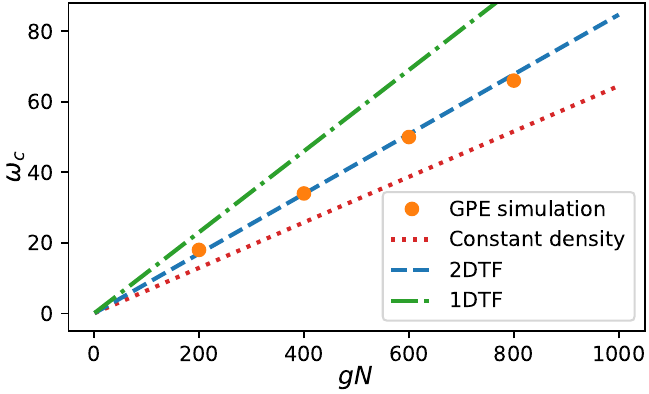}
    \caption{Critical \red{stabilization} anisotropy ratio $\omega_c$ as a function of the parameter $gN$. Orange dots indicate the values obtained from numerical GP simulations. The numerical uncertainties are smaller than the marker size for all simulation points. Three theoretical predictions are shown for comparison: the constant-density approximation (dotted), the 2DTF model (dashed), and the 1DTF model (dash-dotted). The constant-density model slightly underestimates the critical \red{stabilization} anisotropy, and the 1DTF model appears to overestimate it, whereas the 2DTF model provides an excellent fit to the numerical data.}
    \label{fig:omega_c-vs-N}
\end{figure}

Applying this stability condition across a range of parameters, we extract the critical \red{stabilization} anisotropy $ \omega_c $ as a function of interaction strength and particle number. Remarkably, we find that the resulting critical line, \red{as illustrated in Figure~\ref{fig:omega_c-vs-N},} is well approximated by
\begin{equation}
\omega_c^{\text{(num)}} = 0.0833\, g N,
\end{equation}
which agrees  closely with our analytical prediction $ \omega_c \simeq 0.0847\, g N $ derived from the variational model with a TF background. The numerical and analytical values are so close that the match is likely somewhat fortuitous; however, it reinforces the validity of the underlying physical picture.

\section{The Most Unstable Mode \label{sec:most}}

The variational approach developed in the previous section provides valuable insight into the dynamical stability of a dark soliton in an anisotropic trap. However, it also exhibits two notable shortcomings that we now aim to address. 

First, the variational ansatz used so far does not include a spatial phase offset in the sinusoidal modulation along the soliton plane—specifically, the term $ \sin(ky) $ in the ansatz~\eqref{eqn:ansatz-crossover} lacks a phase degree of freedom. While this omission is inconsequential in the case of a uniform background density, it becomes relevant in the presence of spatial inhomogeneity, such as the TF profile used here. Inhomogeneous density profiles can couple the amplitude and phase of perturbations in nontrivial ways, and the lack of a phase parameter may artificially constrain the dynamics and stability landscape captured by the ansatz.

Second, the current analysis predicts the existence of unstable modes in the long-wavelength limit (small $ k $) for any finite condensate, even though the background density vanishes outside a finite region. Physically, however, transverse modulations with wavelengths exceeding the size of the condensate should be suppressed due to the lack of supporting density at large $ y $. The presence of persistent long-wavelength instability despite the finite extent of the condensate is thus an unphysical artifact of the current formulation.

We now construct an improved variational ansatz that includes a transverse phase offset. The modified ansatz is:
\begin{equation}
\psi(x,y,t) = \sqrt{n} \sqrt{1 - \frac{y^2}{R_y^2}} \left( \tanh\left(\frac{x}{\xi_s}\right) + i A(t) \frac{\sin(k y + \phi)}{\cosh\left(x/\xi_s\right)} \right),
\end{equation}
where $ \phi $ is a constant phase shift in the transverse modulation, treated as a fixed parameter. The remaining notation is as in the previous section, with $ A(t) = \alpha(t) + i\beta(t) $ the complex-valued perturbation amplitude and $ \xi_s $ the soliton coherence length. The background density profile retains the TF form, and the perturbation remains localized along the soliton plane via the $ \mathrm{sech}(x/\xi_s) $ envelope.

The derivation of the effective Lagrangian and the resulting equations of motion proceeds along the same lines as in the previous section.   However, the presence of the additional phase $ \phi $ modifies the structure of the integrals over the transverse direction. As a result, the oscillatory functions $ F(x) $, $ G_1(x) $, $ G_2(x) $, and $ G_3(x) $ that appear in the effective dynamics are replaced by generalized versions that account for the phase offset.

Specifically, we define the following functions where oscillatory terms change sign:
\begin{align}
\bar{F}(x) &= \frac{4}{3} - \frac{1}{x^2} \cos(2x) + \frac{1}{2x^3} \sin(2x), \\
\bar{G}_1(x) &= \frac{2}{15} - \frac{x^2 - 3}{2x^4} \cos(2x) - \frac{3 - 5x^2}{4x^5} \sin(2x), \\
\bar{G}_2(x) &= \frac{2}{5} - \frac{(3 - 16x^2) \sin(4x)}{512 x^5} + \frac{(3 - 4x^2) \sin(2x)}{4 x^5} \nonumber \\
&\quad - \frac{3 \cos(2x)}{2x^4} + \frac{3 \cos(4x)}{128x^4}, \\
\bar{G}_3(x) &= \frac{8}{15} - \frac{3}{2x^4} \cos(2x) -\frac{4x^2 - 3}{4x^5} \sin(2x).
\end{align}

The phase-dependent Lagrangian is then obtained by a simple weighted average of the original and phase-shifted functions:
\begin{align}
F(x) &\rightarrow \cos^2\phi\, F(x) + \sin^2\phi\, \bar{F}(x), \\
G_{1}(x) &\rightarrow \cos^2\phi\, G_1(x) + \sin^2\phi\, \bar{G}_1(x), \\
G_{2}(x) &\rightarrow \cos^2\phi\, G_2(x) + \sin^2\phi\, \bar{G}_2(x), \\
G_{3}(x) &\rightarrow \cos^2\phi\, G_3(x) + \sin^2\phi\, \bar{G}_3(x).
\end{align}
Consequently, the linearized oscillation mode frequencies are given as
\begin{align}
&\Omega^2(k) = \mu \left( -\frac{1}{6} - \frac{k^2}{2 \mu} \right) \nonumber \\
& \times \left( \frac{5}{6} - \frac{k^2}{2 \mu}- 2 \frac{\cos^2 \phi (G_1+G_3/3)+\sin^2 \phi (\bar{G_1}+\bar{G_3}/3)}{\cos^2(\phi)  F+\sin^2(\phi) \bar{F}}  \right).
\end{align}

We now turn to the task of identifying the most unstable transverse mode, characterized by the modulation phase $ \phi $ and wavevector $ k $, which grows fastest during the evolution of the snake instability. In an experimental setting, such a mode will dominate the early-time dynamics of the decay, shaping the characteristic patterns observed in the condensate. 

To determine the most unstable configuration, we evaluate the oscillation frequency $ \Omega^2(k) $ as a function of both $ k $ and $ \phi $. 
We find that the most unstable perturbation occurs not for the original
form of the ansatz wavefunction with $ \sin(ky) $ ($ \phi = 0 $), but with $ \cos(ky) $, corresponding to $ \phi = \pi/2 $, which coincide with our numerical simulations of GP equation at low anisotropy, as in Fig.~\ref{fig:SI} (upper middle panel). Moreover, we find that the wavenumber corresponding to the fastest-growing mode is remarkably stable across a broad range of parameters. Unless the system is tuned very close to the stabilization threshold, the most unstable wavevector satisfies
\begin{equation}
k_* R_y \simeq 2.17.
\end{equation}
Using the more unstable cosine mode ($ \phi = \pi/2 $) to estimate the stabilization threshold yields a significantly higher critical \red{stabilization} anisotropy value, $ \omega_c \simeq 0.140\, gN $, than what is observed in numerical simulations. This discrepancy suggests that while the cosine mode dominates in the deeply unstable regime, it does not govern the actual threshold for stability. Indeed, our simulations reveal that as the system approaches stabilization, the snake instability initiates via a sine-like transverse perturbation—odd in $ y $, as can be observed in Fig.~\ref{fig:Ft_plot}. This explains the excellent agreement between the earlier zero-phase ($ \phi = 0 $) calculation and the numerical threshold values. In other words, although the cosine mode is intrinsically more unstable, it becomes energetically suppressed near the boundary of the condensate, allowing the sine mode to dominate close to the transition.

To further explore this behavior, we perform an additional variational calculation assuming a Gaussian profile in the transverse ($ y $) direction, which allows us to explicitly include the kinetic energy cost associated with bending of the background density. We assume a factorized ansatz for the condensate wavefunction, in which the transverse density profile is approximated by a normalized Gaussian of width $ \sigma_y $, while the longitudinal structure $ \Phi(x) $ remains as a variational degree of freedom: 
\begin{equation} \label{eqn:density-Gaussian}
\psi(x,y) = \Phi(x)\, \frac{1}{\sqrt{\sigma_y \sqrt{\pi}}}\, e^{-y^2/(2 \sigma_y^2)}.
\end{equation}
Substituting this ansatz into the GP equation and integrating over the transverse coordinate yields an effective one-dimensional equation for $ \Phi(x) $, with a renormalized interaction strength. The resulting equation is:
\begin{equation}\label{eqn:gpe-gaussian}
-\frac{1}{2} \partial_x^2 \Phi + \frac{1}{2} x^2 \Phi + \frac{g}{\sqrt{2\pi} \sigma_y} |\Phi|^2 \Phi = 
\tilde\mu\Phi,
\end{equation}
where we define the renormalized chemical potential as $ \tilde{\mu} = \mu - \frac{1 + \omega^2 \sigma_y^4}{4 \sigma_y^2} $, which incorporates both the zero-point kinetic energy of the transverse Gaussian and the potential energy associated with the harmonic confinement in the $ y $-direction.

To study the snake instability in this geometry, we now construct a variational ansatz analogous to Eq.~\eqref{eqn:ansatz-crossover}, including transverse modulations and using the same Gaussian envelope. Introducing the dimensionless parameter $ z = k \sigma_y $, we derive the effective Lagrangian as a function of the perturbation amplitude $ A(t) $ and examine the resulting dynamical equations. See the appendix for details.  
The linearized dynamical equations take the form
\begin{align}
\dot{\alpha} &= -  \left[\frac{5}{6} \tilde{\mu} - \frac{\tilde{\mu}}{(1+e^{-z^2/2})}-\frac{z^2 (1+\omega^2 \sigma_y^4 e^{-z^2})}{2\sigma_y^2(1-e^{-z^2})} \right]  \beta, \label{eq:Gauss_lin_alpha} \\
\dot{\beta}  &= + \left[\frac{5}{6} \tilde{\mu} - \frac{\tilde{\mu}}{3(1+e^{-z^2/2})}-\frac{z^2 (1+\omega^2 \sigma_y^4 e^{-z^2})}{2\sigma_y^2(1-e^{-z^2})}\right] \alpha. \label{eq:Gauss_lin_beta}
\end{align}

A key observation is that, unlike in the TF treatment, the above equations admit completely stable solutions for all $ k $, including the low-wavenumber limit. This stabilization arises from the transverse kinetic energy cost introduced by the Gaussian envelope. To investigate this effect more quantitatively, we insert the one-dimensional TF results for a strongly confined system: $ \sigma_y = 1/\sqrt{\omega} $, and 
\begin{equation}
\tilde{\mu} = \left( \frac{3 g N}{8\sqrt{\pi} \sigma_y} \right)^{2/3}.
\end{equation}
Substituting these expressions into Eqs.~\eqref{eq:Gauss_lin_alpha}–\eqref{eq:Gauss_lin_beta} and analyzing the small-$ k $ limit yields a simple condition for complete stabilization: $ \omega \geq \frac{2}{3} \tilde{\mu} $. Using the 1DTF expression for $ \tilde{\mu} $, this translates into a critical \red{stabilization} anisotropy:
\begin{equation} \label{eqn:omega_c-Gaussian}
\omega_c = \left( \frac{2}{3} \right)^{3/2}  \frac{3}{8\sqrt{\pi}}\, g N \simeq 0.115\, g N.
\end{equation}

This value is markedly higher than what we observe in numerical simulations. Furthermore, it is still too low to justify the 1DTF limit used in its derivation. \red{In other words, the critical \red{stabilization} anisotropy in Eq.~\eqref{eqn:omega_c-Gaussian} suggests the stabilization of dark solitons within the 2DTF regime, which is incompatible with the Gaussian form of the background density in Eq.~\eqref{eqn:density-Gaussian}.} In this sense, the calculation is internally inconsistent, as it predicts stabilization in a parameter regime where the assumed Gaussian profile is no longer appropriate. Nonetheless, the calculation provides two important insights. First, it demonstrates that the inclusion of transverse kinetic energy can stabilize all perturbation modes, including those at long wavelengths. Second, it shows that the transverse kinetic cost is not isotropic with respect to perturbation parity: cosine-like modulations incur a larger energy penalty due to their curvature at the edges of the cloud, where the density falls off sharply. This asymmetry is reminiscent of the effective potential barrier experienced by vortices near the condensate boundary, often interpreted in terms of image vortex effects. 

This analysis also reinforces the validity of the TF-based calculation presented in the previous section: although it neglects the transverse kinetic energy, it correctly identifies the sine-like perturbation as the dominant unstable mode near the stabilization threshold. The Gaussian treatment shows that this mode is less affected by edge-induced kinetic energy penalties, explaining why the zero-phase ansatz yields such good agreement with numerical results despite its simplified assumptions.

\section{Conclusion\label{sec:conc}}

We have introduced a variational wavefunction that effectively captures the dynamics of the snake instability of a dark soliton in a trapped BEC. This unified ansatz not only reproduces the bending of the soliton plane and the formation of vortex-antivortex structures, but also provides accurate predictions for the time evolution of stable modes, the critical trap anisotropy required for stabilization, and the wavelength of the most unstable transverse modulation. These predictions are validated by full GP simulations, and the characteristic features we identify—such as the emergence of vortex chains and the stabilization of the soliton at high anisotropies— should be experimentally observable in current cold atom setups. \stfirst{Unlike linearized Bogoliubov analyses, our variational framework extends to nonlinear dynamics and naturally incorporates the effects of inhomogeneous confinement, offering a more comprehensive picture of soliton stability in realistic geometries.}

\red{Notably, our method is numerically more efficient than standard numerical approaches. Simplifying the problem to a small set of coupled Euler-Lagrange equations for variational parameters enables a quick and transparent exploration of the parameter space without the need for repeated BdG diagonalizations or costly full-grid time evolutions. Using a fully analytical approach, we demonstrate unconditional soliton stability without relying on extra assumptions\cite{KevrekidisPRA2004}, a feature that has previously been lacking in the literature. Our ansatz also provides a unified model that captures both the snake instability and the long-lived vortex–soliton oscillations, serving as a practical and extendable tool for studying nonlinear interconversion dynamics. Finally, by focusing on 2D condensates under anisotropic confinement—a regime that has received limited analytical attention—our model directly links experimentally accessible quantities such as the trap aspect ratio and central density.}

Our approach also opens the door to several generalizations. \red{
One possible avenue is to investigate the snake instability of shell solitons \cite{Wang2016PRA}. However, extending the variational framework to three dimensions presents significant challenges due to the additional degrees of freedom and the role of vortex rings in the decay process \cite{Shomroni2009}, which are not easily captured by a low-dimensional ansatz \cite{MithunJPCM2022}.
Multiple interacting solitons \cite{Ake2019PRR}, or dark solitons in exotic trapping potentials suppressing snake instability \cite{Ma2010PRA}, or ring solitons \cite{Theocharis2003PRL, Kevrekidis2017PRL, Toikka2013PRA} would be more tractible through two-dimensional variants of our ansatz.}  A generalization to gray solitons can be similarly considered, where the soliton is no longer stationary but moves at a finite velocity. While the wavefunction structure can be adapted to such a scenario, the resulting dynamics become substantially more complex, as the background motion of the soliton must be disentangled from the instability modes \cite{BuschPRL2000}. Another natural direction is the inclusion of dissipative effects, such as thermal damping or atom loss, which are relevant in experimental settings \cite{FedichevPRA1999}. Since dark solitons correspond to energy maxima, incorporating energy decay mechanisms into the dynamical equations could yield new insights into their long-time evolution and decay pathways \cite{JacksonPRA2007}. Moreover, the emergence of box-trap geometries and the ability to engineer arbitrary potentials in cold atom systems suggest that this variational method could be adapted to study soliton stability in a wider class of confinement landscapes.

Dark solitons are fundamental nonlinear excitations that appear in a wide range of physical systems beyond atomic condensates, including nonlinear optics, water waves, and exciton-polariton fluids. The variational framework developed here may serve as a blueprint for analogous treatments in these contexts, where similar instabilities and confinement effects arise. Moreover, our results suggest a broader program of applying physically motivated variational perturbations to study higher-dimensional nonlinear structures, such as Jones–Roberts solitons \cite{MeyerPRL2017}, which remain less understood despite their experimental relevance. A systematic variational approach may provide new insights into their stability and dynamics under realistic conditions. We hope that our findings not only motivate future experiments probing soliton stability in tunable trapping geometries but also encourage parallel theoretical developments applying variational techniques to a wider class of nonlinear wave phenomena.

\acknowledgements
This work is supported by the Scientific and Technological Research Council of Turkiye (TUBITAK) 1001 program
Project No. 124F122 (AK, MOO). U.T. is partially supported
by the M.Sc. scholarship TUBITAK 2210.
We present the scripts and produced data for this work on Zenodo\cite{zenodo}.

\appendix

\section{}

In the 2DTF approximation, the condensate wavefunction occupies the spatial domain $D = \{(x,y)\in\mathbb R^2,\ |y|<R_y\}$, where $R_y$ is the TF radius in the confined $y$-direction.
The density associated with the ansatz wavefunction \eqref{eqn:ansatz-TF} takes the form
\begin{align} \label{eqn:ansatz-TF-density}
    |\psi|^2 &= n\left(1-\frac{y^2}{R_y^2}\right) \Bigg(\tanh^2(x/\xi_s) +  |A|^2 \frac{\sin^2(k y )}{\cosh^2\left(x/\xi_s\right)} \nonumber \\
    &- 2\mathrm{Im}[A]\frac{\tanh(x/\xi_s)}{\cosh(x/\xi_s)}\sin(ky) \Bigg).
\end{align}
The last term, which is proportional to $\mathrm{Im}[A]$, integrates to zero over the domain $D$ due to its odd symmetry in the $x$-direction. Therefore, the only $A$-dependent term contributing to the Lagrangian $L$ and the variational dynamics is the one proportional to $|A|^2$. The corresponding spatial integral yields a function of the dimensionless parameter $kR_y$:
\begin{align}
    \iint\limits_{D}&\mathrm dx\mathrm dy\ n\left(1-\frac{y^2}{R_y^2}\right) \frac{\sin^2(k y )}{\cosh^2\left(x/\xi_s\right)}  \nonumber \\
    &= \frac{4nR_y\xi_s}{3} + \frac{\xi_sn\cos(2kR_y)}{k^2R_y} -\frac{\xi_sn\sin(2kR_y)}{2k^3R_y^2} \nonumber\\
    &= nR_y\xi_s\left( \frac{4}{3} + \frac{\cos(2kR_y)}{k^2R_y^2} -\frac{\sin(2kR_y)}{2k^3R_y^3} \right) ,
\end{align}
where the term inside the parentheses defines the function $F(kR_y)$ introduced in Eq.~\eqref{eq:F}.

The contribution from the transverse trap potential yields a second integral,

\begin{align}
    \iint\limits_{D}\mathrm dx\mathrm dy\ \frac{\omega^2y^2}{2}n\left(1-\frac{y^2}{R_y^2}\right)  \frac{\sin^2(k y )}{\cosh^2\left(x/\xi_s\right)}  \nonumber \hspace{3.5em}\\
    = n\omega^2R^3\xi_s\Bigg(\frac{2}{15}+\frac{(kR_y)^2-3}{2(kR_y)^4}\cos(2kR_y) \nonumber \\
     \hspace{4.5em}+ \frac{3-5(kR_y)^2}{4(kR_y)^5}\sin(2kR_y)\Bigg),
\end{align}
where the terms inside the parentheses can be written as $G_1(kR_y)$ using the function introduced in Eq.~\eqref{eq:G1}.

The quartic interaction term in the effective potential $U$ \eqref{eqn:U} contributes via the integrals as
\begin{align}
    \iint\limits_{D}\mathrm dx\mathrm dy\ \frac{g}{2}n^2\left(1-\frac{y^2}{R_y^2}\right)^2  \frac{\sin^4(k y )}{\cosh^4\left(x/\xi_s\right)}  \nonumber \hspace{4em}\\
     =  \frac{2n^2gR_y\xi_s}{3}\Bigg(\frac{2}{5} + \frac{(3-16(kR_y)^2)\sin(4kR_y)}{512(kR_y)^5} \nonumber\\
     -\frac{(3-4(kR_y)^2)\sin(2kR_y)}{4(kR_y)^5}  \\
     + \frac{3\cos(2kR_y)}{2(kR_y)^4} -\frac{3\cos(4kR_y)}{128(kR_y)^4}\Bigg),\nonumber
\end{align}
\begin{align}
    \iint\limits_{D}\mathrm dx\mathrm dy\ gn^2\left(1-\frac{y^2}{R_y^2}\right)^2 \frac{\tanh^2(x/\xi_s)}{\cosh^2(x/\xi_s)} \sin^2(ky) \nonumber \\
    =  \frac{2n^2gR_y\xi_s}{3}\Bigg(\frac{8}{15} + \frac{3}{2k^4R_y^4}\cos(2kR_y) \nonumber\\ 
    + \frac{4k^2R_y^2-3}{4k^5R_y^5}\sin(2kR_y)\Bigg),
\end{align}
where the terms inside parentheses can be written in a compact form as $G_2(kR_y)$ and $G_3(kR_y)$ defined in Eqs.\eqref{eq:G2}-\eqref{eq:G3}, respecvively.

In the 1DTF approximation, we assume the condensate has a Gaussian envelope along the strongly confined $y$-direction. Accordingly, we define the variational ansatz wavefunction in the 1DTF  as
\begin{equation} \label{eqn:ansatz-gaussian}
\psi= \frac{\sqrt{n}}{\sqrt[4]{\pi \sigma_y^2}}e^{-\frac{y^2}{2\sigma_y^2}}\Big[\tanh \left({x}/{\xi_s}\right)+\mathrm i A\ \sech(x/\xi_s) \sin(ky)\Big],
\end{equation}
where $\sigma_y$ is the Gaussian width in the confined direction and $n$ is the constant parameter that controls the peak density.

We calculate the Lagrangian $L$ defined in Eq.~\eqref{eqn:L} by substituting this ansatz and integrating over the entire two-dimensional space $\mathbb R^2$. Scaling the Lagrangian with $\left(n\xi_s(1-e^{-z^2})\right)^{-1}$, where $z=k\sigma_y$, the result becomes:
\begin{equation}
    \begin{split}
        L = \beta\dot\alpha-\alpha\dot\beta + \left(\frac{5\tilde\mu}{6} - \frac{z^2}{2\sigma_y^2}\frac{1+\omega^2\sigma_y^4e^{-z^2}}{1-e^{-z^2}}\right)(\alpha^2+\beta^2) \\
        -\frac{\tilde \mu }{3}   \frac{1}{1+e^{-z^2}} (\alpha^2+3\beta^2) \\
        - \frac{\tilde \mu}{12} \frac{3-4e^{-z^2/2}+e^{-2z^2}}{1-e^{-z^2}} ((\alpha^2+\beta^2))^2.
    \end{split}
\end{equation}
\bibliography{ref}

\end{document}